\documentclass[twocolumn,seceq]{jpsj2}

\def\runtitle{
Electronic and Lattice Dynamics in The Photoinduced I-to-N Phase Transition.
}
\def\runauthor{
Naoyuki \textsc{Miyashita}, Makoto \textsc{Kuwabara} and Kenji \textsc{Yonemitsu}
}

\title{
Electronic and Lattice Dynamics in The Photoinduced Ionic-to-Neutral Phase Transition in a One-Dimensional Extended Peierls-Hubbard Model
}

\author{
Naoyuki \textsc{Miyashita}$^{1}$
\thanks{E-mail address: miya@ims.ac.jp},
Makoto \textsc{Kuwabara}$^{2}$ and
Kenji \textsc{Yonemitsu}$^{1,3}$ }

\inst{
$^1$Graduate University for Advanced Studies, 
Okazaki 444-8585 \\
$^2$Molecular Photoscience Research Center, 
Kobe University, Nada, Kobe 657-8501 \\
$^3$Institute for Molecular Science, 
Okazaki 444-8585
}

\recdate{\today}

\abst{
Real-time dynamics of charge density and lattice displacements is studied during photoinduced ionic-to-neutral phase transitions by using a one-dimensional extended Peierls-Hubbard model with alternating potentials for the one-dimensional mixed-stack charge-transfer complex, TTF-CA.
The time-dependent Schr\"odinger equation and the classical equation of motion are solved for the electronic and lattice parts, respectively. 
We show how neutral domains grow in the ionic background. 
As the photoexcitation becomes intense, more neutral domains are created.
Above threshold intensity, the neutral phase is finally achieved.
After the photoexcitation, ionic domains with wrong polarization also appear.
They quickly reduce the averaged staggered lattice displacement, compared with the averaged ionicity.
As the degree of initial lattice disorder increases, more solitons appear between these ionic domains with different polarizations, which obstruct the growth of neutral domains and slow down the transition.
}

\kword{
photoinduced phase transition, dynamics, domain wall,
time-dependent Schr\"odinger equation, TTF-CA, charge-transfer complex, one-dimensional extended Peierls-Hubbard model
}

\begin{document}
\sloppy
\maketitle

\section{Introduction}


Dynamics in non-equilibrium phase transitions has been challenging and studied from different viewpoints.
In cooperative processes, roles played by microscopic degrees of freedom are interesting.
When photoirradiation triggers a phase transition, induced real-time dynamics gives important information on their roles.
We here focus on photoinduced phase transitions in quasi-one-dimensional mixed-stack charge-transfer (CT) complexes, which have been studied extensively both experimentally\cite{koshihara90,koshihara95,koshihara99,suzuki99,koshihara00, tanimura01,iwai02,luty02} and theoretically\cite{nasu-ed,huai00,iwano02}.

Quasi-one-dimensional mixed-stack CT complexes consist of donor (D) and acceptor (A) molecules, which are alternately stacked to form columns.
They exhibit a neutral-ionic phase transition in equilibrium by applying hydrostatic pressure or by lowering the temperature\cite{torrance81a}.
In these complexes, electrons are itinerant and an electron-lattice coupling plays a key role in causing the phase transition.
At the phase transition charge is transferred between D and A molecules, accompanied by a distortion forming (D$^{+\delta}$ A$^{-\delta}$) dimers.
The two phases are distinguished from each other by the ionicity ($\bar{\rho}$). It is a measure of the charge transfer and defined later so that the neutral (N) phase has $\bar{\rho} < 0.5$, while the ionic (I) phase has $\bar{\rho} > 0.5$.

Tetrathiafulvalene-{\it p}-chloranil (TTF-CA) is one of such CT complexes and shows a photoinduced phase transition\cite{koshihara90,koshihara95,koshihara99,suzuki99,koshihara00, tanimura01,iwai02,luty02}. 
In the TTF-CA complex, the TTF molecule is a donor, and the CA molecule is an acceptor\cite{torrance81}.
At ambient pressure, it is ionic below $T_{\mathrm N \mathrm I}=81$ K and neutral above $T_{\mathrm N \mathrm I}$.
In the ionic phase, two structural phases exist.
When the inversion symmetry is lost, it becomes ferroelectric\cite{MLCointe95}.
It is realized at ambient pressure in equilibrium. 
When the polarizations D$^{+\delta}$A$^{-\delta}$ and A$^{-\delta}$D$^{+\delta}$ are disordered to recover the symmetry, it becomes paraelectric.
It is realized under high pressure\cite{MHLemee-C97} or just after the photoexcitation\cite{luty02}.
The pressure-temperature phase diagram is similar to typical solid-liquid-gas phase diagrams.

Recently, the time-resolved reflectivity has been measured with improved resolution, which reflects the evolution of the spatially averaged ionicity\cite{suzuki99,tanimura01,iwai02}.
Time-resolved second harmonic generation (SHG) has also been measured to study the evolution of the degree of the inversion-symmetry breaking after photoexcitations\cite{luty02}.
The SHG intensity decreases rapidly after the photoirradiation of the ferroelectric ionic phase, but the ionicity decreases much slower.
This information is important to reveal non-equilibrium processes in the photoinduced phase transition.
Because the reflectivity and the SHG intensity reflect the averaged properties, we do not know creation or annihilation processes of neutral and ionic domains.
Space-resolved properties of the coupled charge-lattice system are important to clarify the non-equilibrium processes during the photoinduced phase transition.

Theoretically, Koshino and Ogawa have studied photoinduced structural dynamics in a one-dimensional system composed of classical lattice displacements and localized electrons occupying one of the two levels per site\cite{koshino98}.
The dynamics is described by nucleation of a kink-antikink pair and motion of the kinks.
The strength and the range of the elastic-lattice coupling have been shown to decide whether a global change is realized or not.
Meanwhile, Huai {\it et al.} have studied the stability of a neutral domain in the ionic background in a quasi-one-dimensional itinerant-electron-lattice model for TTF-CA\cite{huai00}.
The origin of threshold excitation intensity for the transition is discussed, based on the adiabatic potential as a function of the domain size.
Its dynamics was not studied there.
Only few theoretical studies have so far been devoted to dynamics in photoinduced phase transitions of itinerant-electron-lattice systems\cite{iwano02,mya02a,mya02b}.
Our previous and preliminary studies have shown only that the motion of the ionicity has three characteristic time scales, which originate from the electronic motion, the lattice motion, and the neutral-ionic domain-wall motion, during the ionic-to-neutral phase transition\cite{mya02a,mya02b}.

In this paper, we study the dynamics of coupled charge density and lattice displacements during the photoinduced ionic-to-neutral phase transition in a one-dimensional extended Peierls-Hubbard model with alternating potentials for TTF-CA.
We show how neutral domains grow or shrink in the ionic background.
The dynamics is described by domain walls, which exist between ionic and neutral domains or between ionic domains with different polarizations.
In \S\ref{model}, the model and the method are described.
In \S\ref{results}, results are presented. Finally in \S\ref{conc}, the conclusion is given.

\section{Extended Peierls-Hubbard Model with Alternating Potentials}\label{model}
	
The highest occupied molecular orbital (HOMO) of the D molecule and the lowest unoccupied molecular orbital (LUMO) of the A molecule are necessary to describe the electronic state. 
In the limit of vanishing transfer integrals $t_0 \to 0$, the HOMO is doubly occupied in the neutral phase, while both of the HOMO and the LUMO are singly occupied in the ionic phase.
Different types of electron-lattice couplings have so far been employed to explain the dimerization\cite{nagaosa86-3,painelli02,kxy02,sakano96}.
Here we assume that the Coulomb interaction strength is modified by the lattice displacement\cite{sakano96}.
We use a one-dimensional extended Peierls-Hubbard model with alternating potentials at half filling, which is a single-chain analogue of the model used in ref.~\citen{huai00},
\begin{equation}
H =  H_\mathrm{el} + H_\mathrm{lat} \;,\label{g-ham} 
\end{equation}
with
\begin{align}
H_\mathrm{el} = & 
-t_0 \sum _{\sigma,l=1}^{N}
   \left( c^{\dagger }_{l+1,\sigma}c_{l,\sigma}
     + \mathrm{h.c.} \right)  \nonumber \\ 
& +\sum _{l=1}^{N} 
\left[ U n_{l,\uparrow} n_{l,\downarrow} 
+ (-1)^{l} \frac{d}{2} n_{l} \right] \nonumber \\ 
& +\sum _{l:odd}^{N} 
\bar{V}_{l} (n_{l}-2) n_{l+1}
+ \sum _{l:even}^{N} 
\bar{V}_{l} n_{l} (n_{l+1}-2) \;, \label{e-ham} \\
H_\mathrm{lat} =& \sum _{l=1}^{N}
\left[ \frac{k_1}{2}y_{l}^{2}
+\frac{k_2}{4}y_{l}^{4}
+\frac{1}{2}m_{l}\dot{u}_{l}^{2} \right] \;,\label{l-ham} 
\end{align}
where, $ c^{\dagger }_{l,\sigma} $  ($ c_{l,\sigma} $) is the creation (annihilation) operator of a $\pi$-electron with spin $\sigma$ at site $l$, $ n_{l,\sigma} = c^{\dagger}_{l,\sigma} c_{l,\sigma} $, $ n_{l} = n_{l,\uparrow} + n_{l,\downarrow} $, $ u_{l} $ is the dimensionless lattice displacement of the $l$th molecule along the chain from its equidistant position, and $ y_{l} = u_{l+1} - u_{l} $.
The distance between the $l$th and $(l+1)$th molecules is then given by $r_l=r_0 (1+u_{l+1}-u_l)$, where $r_0$ is the averaged distance between the neighboring molecules along the chain.
The D and A molecules are located at the odd and even sites, respectively.
The nearest-neighbor repulsion strength between the $ l $th and $ (l+1) $th sites $ \bar{V}_{l} $ depends on the bond length $ y_{l} $, $ \bar{V}_{l} = V + \beta _{2} y^{2}_{l} $, where $ V $ is for the regular lattice, and $\beta _{2}$ is the quadratic coefficient. 
The linear coefficient is neglected for simplicity.
The parameter $ t_0 $ denotes the nearest-neighbor transfer integral, $ U $ the on-site repulsion strength, and $ d $ the level difference between the HOMO and the LUMO in the neutral phase with $t_0=0$. 
The elastic energy is expanded up to the fourth order: the parameters $ k_{1} $ and $ k_{2} $ are the linear and nonlinear elastic constants. 
The mass of the $ l $th molecule is denoted by $ m_l $. 
The number of sites are denoted by $N$.

\subsection{Hartree-Fock approximation}\label{HFA}

We use the unrestricted Hartree-Fock (HF) approximation, which replaces the electronic part (\ref{e-ham}) by
\begin{equation}
\begin{split}
H^\mathrm{HF}_\mathrm{el}=& 
\sum ^{N}_{\sigma ,l=1}
\left[ h^\mathrm{H}_{l,\sigma}c^{\dagger}_{l,\sigma} c_{l,\sigma}
+h^\mathrm{F}_{l,\sigma}(c^{\dagger}_{l+1,\sigma} c_{l,\sigma}
+c^{\dagger}_{l,\sigma}c_{l+1,\sigma})\right] \\
&-h_{\mathrm c} \;,
\end{split}
\end{equation}
with
\begin{align}
{h}^\mathrm{H}_{l,\sigma} = & U \langle n_{l,\bar{\sigma}} \rangle
+ (-1)^{l} \frac{d}{2} 
-2\delta_{l,even}(\bar{V}_{l}+\bar{V}_{l-1})\nonumber \\ 
   & +\bar{V}_{l-1} \langle n_{l-1} \rangle 
+\bar{V}_{l} \langle n_{l+1} \rangle \;,\\
{h}^\mathrm{F}_{l,\sigma} =&  -\left[ t_{0}
+\bar{V}_{l} \langle c^{\dagger}_{l,\sigma}c_{l+1,\sigma} \rangle \right] \;,\\
{h}_\mathrm{c}  =&  \sum _{l=1}^{N}
\Biggl[ U \langle n_{l,\uparrow} \rangle \langle n_{l,\downarrow}\rangle
+\bar{V}_{l} \biggl( \langle n_{l} \rangle \langle n_{l+1} \rangle \nonumber \\ 
&-\sum _{\sigma} \langle c^{\dagger}_{l,\sigma}c_{l+1,\sigma} \rangle
\langle c^{\dagger}_{l+1,\sigma}c_{l,\sigma} \rangle \biggr) \Biggr] \;, 
\end{align}
where ${h}^\mathrm{H}_{l,\sigma}$ is the Hartree term, 
${h}^\mathrm{F}_{l,\sigma}$ is the Fock term, $\delta_{l,even}=1$ for 
even $l$, and $\delta_{l,even}=0$ otherwise.
We iteratively solve the eigenvalue equation to obtain the ground state that is self-consistent with the one-body densities $\langle c^\dagger_{l,\sigma}c_{l,\sigma} \rangle$ and $\langle c^\dagger_{l,\sigma} c_{l+1,\sigma} \rangle$ and the lattice displacements $y_l$, which satisfy the Hellman-Feynman theorem.

We then add random numbers to the initial $ y_{l} $ and $ \dot{u}_l $  values according to the Boltzmann distribution at a fictitious temperature $\it{T}$.
They roughly correspond to thermal lattice fluctuations just before the photoexcitation.
The degree of initial lattice disorder $\it{T}$ used here is much smaller than the energy barrier between the neutral and ionic phases.
With thus modified lattice displacements, the HF Hamiltonian is diagonalized once again by imposing the self-consistency on the one-body densities only.
Photoexcitations are introduced by changing the occupancy of the highest occupied and lowest unoccupied HF orbitals.

\subsection{Equations of motion}\label{TDE}

For the electronic part, we solve the time-dependent Schr\"odinger equation, 
\begin{align}
\vert \psi_{\nu, \sigma}(t) \rangle = &
\mathrm{T} \exp \left[ -\frac{\mathrm{i}}{\hbar} 
\int_0^t \, {\mathrm d} t^\prime H^\mathrm{HF}_\mathrm{el}(t^\prime) \right]
\vert \psi_{\nu, \sigma}(0) \rangle \; ,\label{tde1}
\end{align}
where $\mathrm{T}$ represents the time-ordering operator.
For $t_j=j \times \Delta t < t < t_j+\Delta t$, it is rewritten as
\begin{align}
\vert \psi_{\nu, \sigma}(t_j + \Delta t) \rangle = &
\mathrm{T} \exp \left[ -\frac{\mathrm{i}}{\hbar} 
\int_{t_j}^{t_j+\Delta t}  {\mathrm d} t^\prime H^\mathrm{HF}_\mathrm{el}(t^\prime) \right]
\vert \psi_{\nu, \sigma}(t_j) \rangle \; .
\end{align}
The exponential operator is decomposed\cite{suzumasu93}, in such a way that the decomposition is accurate to the order of $(\Delta t )^2 $\cite{ono90}.

For the lattice part, we solve the classical equation of motion,
\begin{align}
m_l \frac{{\mathrm d}^2 u_l}{{\mathrm d} t^2}=
-\frac{\partial}{\partial u_l} 
\left( H_\mathrm{lat}(u_l) + \langle 
H_\mathrm{el}(u_l) \rangle \right)\;,
\end{align}
where the expectation value is taken with respect to the wave function (\ref{tde1}).
We employ the leap frog method,
\begin{align}
u_l(t+\Delta t) 
=& u_l(t)+\Delta t v_l(t+\frac{\Delta t}{2})+{\cal O} (\Delta t^3) \; , \\
v_l ( t+\frac{\Delta t}{2} ) 
=& v_l (t-\frac{\Delta t}{2})+\Delta t \frac{F_l (t)}{m_l} + {\cal O} (\Delta t^3) \; , \\
F_l(t) 
=& 
-( 
\frac{\partial H_\mathrm{lat} }{\partial u_l} 
+ \langle \psi_{\nu,\sigma}(t) \vert 
\frac{ \partial H_\mathrm{el} }{\partial u_l} 
\vert \psi_{\nu,\sigma}(t) \rangle ) \; ,
\end{align}
which are accurate to the order of $(\Delta t)^2$. The computation time is shorter than the similarly accurate, velocity Verlet method\cite{vvm82}.
The lattice velocity at time $t$ can be defined as
\begin{align}
v_l(t)=\frac{ v_l(t+\frac{\Delta t}{2}) + v_l(t-\frac{\Delta t}{2}) }{ 2 } \;.
\end{align}
During the numerical calculations we have confirmed that the total energy is always conserved.

\section{Results and Discussions}\label{results}

We use $N$=100, $ t_0 $=0.17 eV, $ U $=1.528 eV, $ V $=0.604 eV, $ d $=2.716 eV, $ \beta_2 $=8.54 eV, $ k_1 $=4.86 eV, $ k_2 $=3400 eV, and the bare phonon energy $\omega \equiv (1/r_0)(2 k_1/m_r)^{1/2} $=0.0192 eV, and impose the periodic boundary condition.
Here the reduced mass $m_r$ is defined as $m_r=m_\mathrm{D}m_\mathrm{A}/(m_\mathrm{D}+m_\mathrm{A})$ with the mass of the donor molecule $m_\mathrm{D}$ and the mass of the acceptor molecule $m_\mathrm{A}$. 
With these parameters, the dimerized ionic phase is stable and the neutral phase is metastable.
These parameters are almost the same as in ref.~\citen{huai00}, but the bare phonon energy $\omega$ used here is about four times higher than the optical phonon energy of the TTF-CA complex in the neutral phase.
Note that $u_l$ here is in the unit of the DA distance in the regular neutral phase.
The averaged ionicity is defined as $\bar{\rho}= 1 + 1/N  \sum_{l=1}^{N} (-1)^{l} \langle n_{l} \rangle $.

\subsection{Adiabatic potentials}\label{pe}

\begin{figure}
\includegraphics[height=5cm]{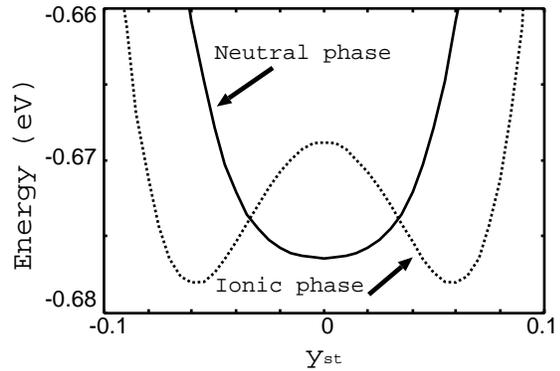}
\caption{Adiabatic potentials in the neutral phase (solid line) and in the ionic phase (dashed line), as a function of the staggered lattice displacement $y_{st}$. 
The parameters are $t_0=$0.17 eV, $U=$1.528 eV, $V=$0.604 eV, $d=$2.716 eV, $\beta_2=$8.54 eV, $k_1=$4.86 eV, and $k_2=$3400 eV.}
\label{pot}
\end{figure}

Figure \ref{pot} shows the adiabatic potentials in the ionic phase (dotted line) and in the neutral phase (solid line), as a function of the staggered lattice displacement $y_{st}$, which is defined as $y_l=(-1)^l y_{st}$.
The molecules are equidistant for $y_{st}=0$, and dimerized for $\vert y_{st} \vert \ne 0$.
The neutral phase has a local minimum at $y_{st}=0$, while the ionic phase has two minima at about $y_{st}= \pm 0.06$.
In the ionic phase, the donor and acceptor molecules have positive and negative charge, respectively.
In the dimerized lattice, the polarization vectors are uniform so that this phase is ferroelectric. 
With the parameters above, the ferroelectric ionic phase is stable, and the neutral phase is metastable.
The initial state in the later calculations is always in the ferroelectric ionic phase.
The optical phonon energy in the ionic phase is 2.4 times larger than that in the neutral phase.

\subsection{Emergence of a few I-\=I solitons after weak photoexcitations.} \label{sol}

\begin{figure}
\includegraphics[height=12cm]{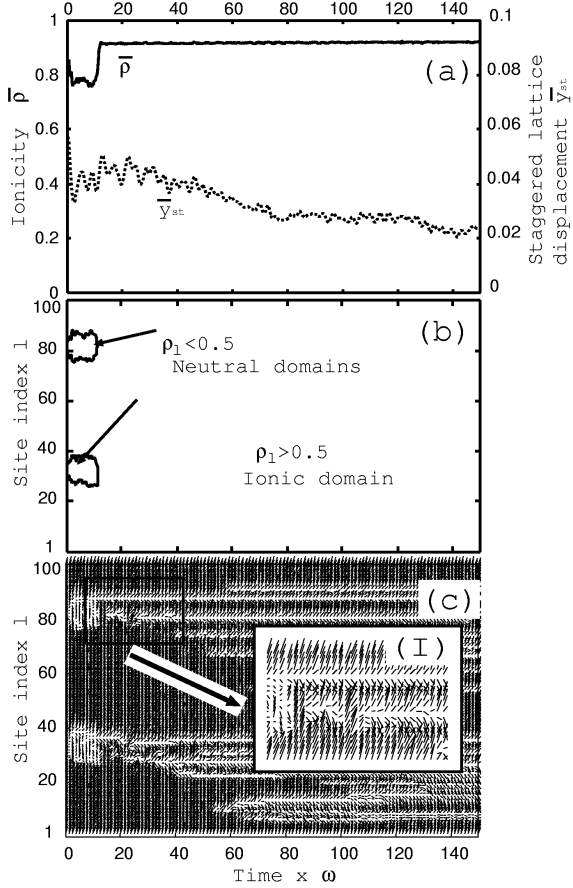}
\caption{(a) Averaged ionicity $\bar{\rho}$ (solid line) and averaged staggered lattice displacement $\bar{y}_{st}$ (dashed line), as a function of time $t$ after the photoexcitation, which is multiplied by the bare phonon energy $\omega$.
(b) Space and time dependence of the boundaries between the neutral and ionic domains ($\rho_l=0.5$).
(c) Correlation between the staggered lattice displacement $y_{st \; l}$ and the ionicity $\rho_l$, as a function of the site index $l$ and the elapsing time $t$ multiplied by $\omega$.
Inset (I) of (c) shows the area indicated by the box on the enlarged scale.
The parameters are $t_0=$0.17 eV, $U=$1.528 eV, $V=$0.604 eV, $d=$2.716 eV, $\beta_2=$8.54 eV, $k_1=$4.86 eV, $k_2=$3400 eV, $\omega=$0.0192 eV, and $T=$0.00017 eV. In the 100-site chain, 2 electrons are excited.}
\label{fig-tl-2ex}
\end{figure}

We show here what happens in the ionic phase after weak photoexcitations.
When two electrons are excited and the degree of initial lattice disorder is very small, $\it{T}$=0.00017 eV, the ionicity and the staggered lattice displacements evolve as shown in Fig.~\ref{fig-tl-2ex}.
The horizontal axis gives the elapsing time $t$ multiplied by $\omega$.
The solid line in Fig.~\ref{fig-tl-2ex}~(a) shows the averaged ionicity $\bar{\rho}$. 
After the photoexcitation, it first decreases from its initial value $\bar{\rho}=0.92$ to $\bar{\rho}=0.76$ and it soon comes back to the initial value by the time about $10/\omega$.
The space and time dependence of the boundaries between the ionic and neutral domains, where $\rho_l=0.5$, is shown in Fig.~\ref{fig-tl-2ex}~(b). 
Here the ionicity $\rho_l$ is defined as $\rho_l = \frac{(-1)^l}{4}(-\langle n_{l-1}\rangle + 2 \langle n_l \rangle - \langle n_{l+1} \rangle ) + 1$.
The vertical axis gives the site index $l$.
It is clearly seen that neutral domains are created immediately after the photoexcitation and they soon disappear by the time about $10/\omega$, which coincides with the behavior of $\bar{\rho}$.
The dotted line in Fig.~\ref{fig-tl-2ex}~(a) shows the magnitude of the averaged staggered lattice displacement defined as $\bar{y}_{st}=1/N\vert \sum_{l=1}^{N}(-1)^l y_l \vert $.
It would roughly be proportional to the SHG intensity.
Its initial value is $\bar{y}_{st}=0.06$.
At first, $\bar{y}_{st}$ quickly drops to $\bar{y}_{st}=0.035$ much before the recovery of $\bar{\rho}$ at the time about 10$/\omega$.
Then, $\bar{y}_{st}$ slightly increases on average to $\bar{y}_{st}=0.045$ by the time about 20$/\omega$.
Finally, $\bar{y}_{st}$ gradually decays after that.
The bars in Fig.~\ref{fig-tl-2ex}~(c) show the correlation between the staggered lattice displacement defined as $y_{st \; l}=\frac{(-1)^l}{4}(-y_{l-1}+2 y_l - y_{l+1})$ and the ionicity $\rho_l$, as a function of the site index $l$ (vertical axis) and the elapsing time $t$ multiplied by $\omega$ (horizontal axis).
\begin{figure}
\includegraphics[height=4cm]{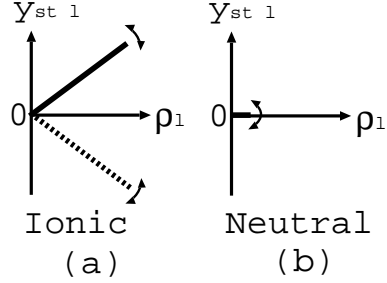}
\caption{Long and slanted bar (a) on distorted ionic sites, and short and horizontal bar (b) on undistorted neutral sites, in Fig.~\ref{fig-tl-2ex}~(c). The vertical component shows $y_{st \; l}$, while the horizontal one $\rho_l$.}
\label{fig-bar}
\end{figure}
As shown in Fig.~\ref{fig-bar}, each bar has two components. 
The vertical component shows $y_{st \; l}$, while the horizontal one $\rho_l$.
Then, distorted ionic sites are described by long and slanted bars [Fig.~\ref{fig-bar}~(a)], while undistorted neutral sites by short and horizontal bars [Fig.~\ref{fig-bar}~(b)].
In ionic domains with (D$^+$A$^-$) [(A$^-$D$^+$)] dimers, $y_{st \; l}$ is positive (negative), so that the bar points up (down) [Fig.~\ref{fig-bar}~(a), solid (dotted) bar].
These domains are denoted by I and \=I domains, respectively, hereafter.
It is clearly seen that, after the photoexcitation, neutral domains with almost equidistant molecules are quickly created at first.
This is why $\bar{y}_{st}$ decays in the corresponding period in Fig.~\ref{fig-tl-2ex}~(a).
Then, small \=I domains appear after the neutral domains disappear [Fig.~\ref{fig-tl-2ex}~(c)], so that $\bar{y}_{st}$ is not recovered to the initial value any more.
\begin{figure}
\includegraphics[height=5cm]{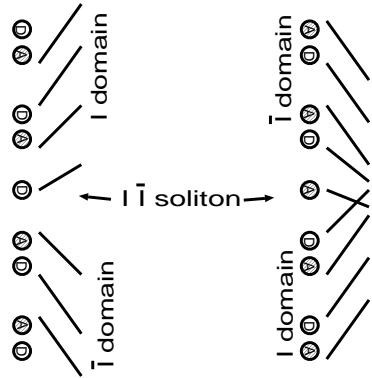}
\caption{I-\=I solitons between I and \=I domains with different polarizations, which are denoted by the directions of the bars. D and A are donor and acceptor molecules, respectively.}
\label{fig-sol}
\end{figure}
The boundary between the I and \=I domains is regarded as a I-\=I soliton, which looks like Fig.~\ref{fig-sol} in Fig.~\ref{fig-tl-2ex}~(c).
Finally, the number of I-\=I solitons gradually increases.
The excitation energy is partially absorbed into these solitons.
In this subsection, we show the creation of small neutral domains by weak photoexcitations and their decay into I-\=I solitons.

\subsection{ Photoinduced ionic-to-neutral phase transition} \label{in-lt}

\begin{figure}
\includegraphics[height=12cm]{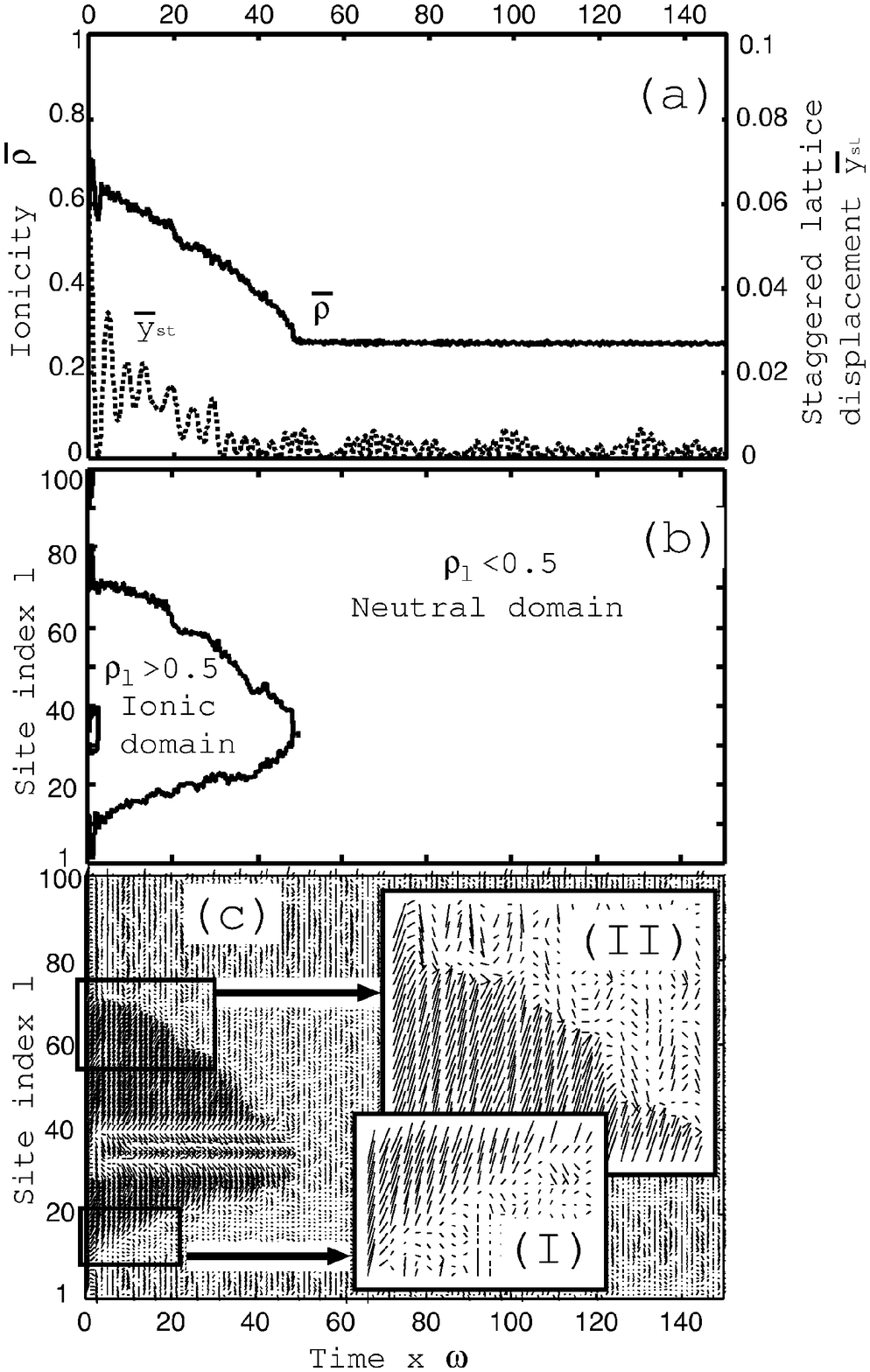}
\caption{(a) Averaged ionicity $\bar{\rho}$ (solid line) and averaged staggered lattice displacement $\bar{y}_{st}$ (dashed line), as a function of time $t$ after the photoexcitation, which is multiplied by the bare phonon energy $\omega$.
(b) Space and time dependence of the boundaries between the neutral and ionic domains ($\rho_l=0.5$).
(c) Correlation between the staggered lattice displacement $y_{st \; l}$ and the ionicity $\rho_l$, as a function of $l$ and $\omega t$.
Insets (I) and (II) of (c) show the areas indicated by the boxes on the enlarged scale.
The parameters are the same as in Fig.~\ref{fig-tl-2ex}.
In the 100-site chain, 4 electrons are excited.}
\label{fig-tl}
\end{figure}

In this subsection, we show how neutral domains grow in the ionic background after strong photoexcitations.
When four electrons are excited and the degree of initial lattice disorder is still very small, $\it{T}$=0.00017 eV, the ionicity and the staggered lattice displacements evolve as shown in Fig.~\ref{fig-tl}.
The seed used to produce random numbers is the same as that in the previous subsection and will be used later throughout the paper.
The meanings of the lines are also the same as before.
After the photoexcitation, the averaged ionicity $\bar{\rho}$ first shows a dip and gradually decreases on average to $\bar{\rho}=0.27$ by the time about $45/\omega$ [Fig.~\ref{fig-tl}~(a), solid line].
The magnitude of the averaged staggered lattice displacement $\bar{y}_{st}$ decreases on average from its initial value $\bar{y}_{st}=0.06$ to $\bar{y}_{st} \sim 0$ by the time about $30/\omega$ [Fig.~\ref{fig-tl}~(a), dotted line].
Two neutral domains are created by the photoexcitation, but the smaller one immediately disappears [Fig.~\ref{fig-tl}~(b)].
Meanwhile, the larger neutral domain expands and annihilates the initial ionic domain by the time about $45/\omega$.
It is clearly seen in Fig.~\ref{fig-tl}~(c) that, when the smaller neutral domain disappears, I and \=I domains appear instead.
Because of the appearance of the \=I domains, the averaged staggered lattice displacement $\bar{y}_{st}$ decreases much faster than the averaged ionicity $\bar{\rho}$.
The expansion of the larger neutral domain is shown in the insets (I) and (II) of Fig.~\ref{fig-tl}~(c) on the enlarged scale.
In the inset (I), the expansion is smooth because both $y_{st \; l}$ and $\rho_l$ change gradually.
In the inset (II), the expansion is stepwise due to I-\=I solitons.
\begin{figure}
\includegraphics[height=5cm]{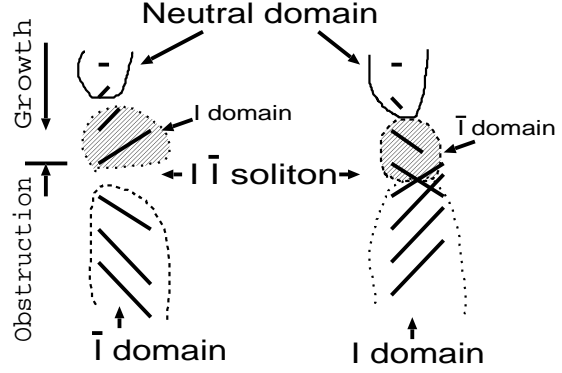}
\caption{I-\=I solitons near the neutral-ionic domain boundary obstruct the growth of the neutral domain.}
\label{solc}
\end{figure}
The I-\=I solitons near the neutral-ionic domain boundary obstruct the growth of the neutral domain, as shown in Fig.~\ref{solc}.
When the small but finite staggered displacements around the neutral-ionic domain boundary become parallel to those of the ionic domain, the neutral domain rapidly expands.
In this subsection, we show the obstruction of the growth of the neutral domain by I-\=I solitons.

\subsection{Case of many I-\=I solitons} \label{rin}

\begin{figure}
\includegraphics[height=12cm]{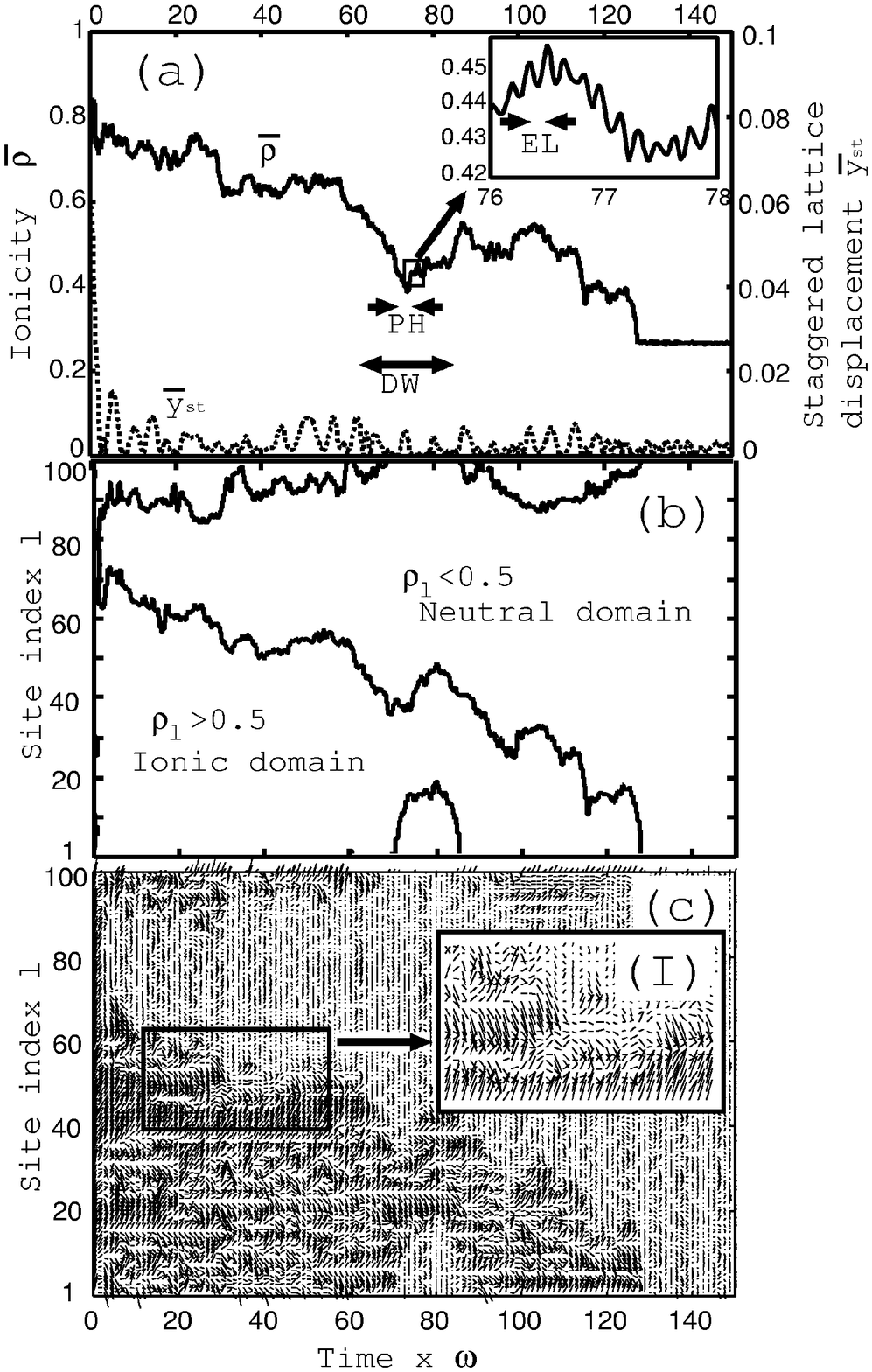}
\caption{(a) Averaged ionicity $\bar{\rho}$ (solid line) and averaged staggered lattice displacement $\bar{y}_{st}$ (dashed line), as a function of time $t$ after the photoexcitation, which is multiplied by the bare phonon energy $\omega$.
(b) Space and time dependence of the boundaries between the neutral and ionic domains ($\rho_l=0.5$).
(c) Correlation between the staggered lattice displacement $y_{st \; l}$ and the ionicity $\rho_l$, as a function of $l$ and $\omega t$.
Inset (I) of (c) shows the area indicated by the box on the enlarged scale.
The degree of initial lattice disorder is $T=$0.0017 eV. The other parameters are the same as in Fig.~\ref{fig-tl-2ex}.
In the 100-site chain, 4 electrons are excited.}
\label{fig-in}
\end{figure}

In this subsection, we show a case of many I-\=I solitons created by the photoexcitation before the ionic-to-neutral phase transition.
The number of I-\=I solitons sensitively depends on of the degree of initial lattice disorder $\it{T}$: it increases with $\it{T}$.
When four electrons are excited and the degree of initial lattice disorder is not very small, $\it{T}$=0.0017 eV, the ionicity and the staggered displacements evolve as shown in Fig.~\ref{fig-in}.
The averaged ionicity decreases from its initial value $\bar{\rho} \sim 0.878$ to $\bar{\rho} \sim 0.266 $ by the time about 130$/\omega$ [Fig.~\ref{fig-in}~(a)]. 
Here three characteristic time scales are observed.
The averaged ionicity $\bar{\rho}$ rapidly oscillates.
We calculated the charge-charge and current-current correlation functions in the random phase approximation (RPA) \cite{kxy93}. 
Their spectra have peaks at about $6\times 2\pi \omega$ and $5\times 2 \pi \omega$, respectively, in both the neutral and the ionic phases at $T=0$.
The corresponding periods are then about $0.16/\omega$ and $0.2/\omega$, respectively.
A period of about $0.16/\omega$ is indicated by ``EL'' in the inset of Fig.~\ref{fig-in}~(a).
The clearly seen beat is due to the oscillation of period about $0.16/\omega$ from the charge-charge correlation and that of period about $2/\omega$ from the current-current correlation.
The slow oscillation of $\bar{\rho}$ indicated by ``PH'' in Fig.~\ref{fig-in}~(a) is correlated with that of $\bar{y}_{st}$.
Thus, the observed periods ranging between $3/\omega$ and $6/\omega$ are due to the lattice oscillations.
It is noted that the period of the noninteracting lattice oscillation is $2\pi/\omega$.
Recall that the optical phonon energy in the ionic phase is $2.4$ times larger than that in the neutral phase.
Thus, the observed range of the period distribution is reasonable.
The slowest time scale is due to the collective motion of domain walls and indicated by ``DW'' in Fig.~\ref{fig-in}~(a).
A growth of the neutral domain is observed during the period between $0$ and $130/\omega$, accompanied with the shrinking ionic background and with the oscillating domain boundaries [Fig.~\ref{fig-in}~(b)].
The decrease of $\bar{\rho}$ in Fig.~\ref{fig-in}~(a) is due to the growth of the neutral domain in Fig.~\ref{fig-in}~(b).
The averaged staggered lattice displacement $\bar{y}_{st}$ decreases much quicker than $\bar{\rho}$ [Fig.~\ref{fig-in}~(a)].
After the photoexcitation, many \=I domains appear [Fig.~\ref{fig-in}~(c)].
It suggests that the SHG intensity becomes weak very quickly, compared with the decrease of the ionicity, even in the one-dimensional model.
Many I-\=I solitons are observed in the inset (I) of Fig.~\ref{fig-in}~(c), and they indeed slow down the growth of the neutral domain.
This situation is already illustrated in Fig.~\ref{solc}.
The I-\=I solitons are obstacles to the expansion of the neutral domain, and they sensitively affect the collective motion of the neutral-ionic domain walls.

\begin{figure}
\includegraphics[height=10cm]{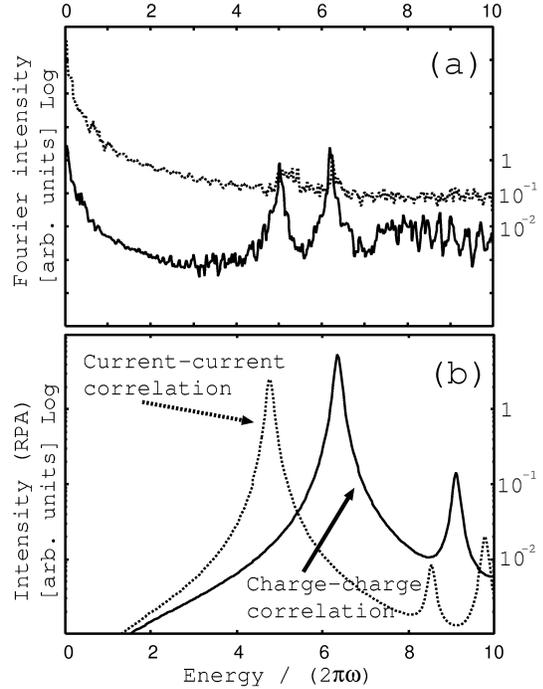}
\caption{(a) Fourier power spectra for the averaged ionicity $\bar{\rho}$ in Fig.~\ref{fig-in}~(a). The dotted line is obtained during the I-to-N transition ($0/\omega < t < 102/\omega$), and the solid line after the transition ($143/\omega < t < 245/\omega$).
(b) Charge-charge (solid line) and current-current (dotted line) excitation spectra in the RPA for the static neutral state.}
\label{fig-fftrpa-in}
\end{figure}

Fourier transforms of some time windows of the ionicity $\bar{\rho}$ in Fig.~\ref{fig-in}~(a) are shown in Fig.~\ref{fig-fftrpa-in}~(a).
Here, the dotted line is obtained from $0  < t < 102/\omega $ during the I-to-N transition, and the solid line from $143/\omega < t < 245/\omega $ after the transition.
During the transition, slow components are dominant and broad reflecting the complex motion of the neutral-ionic domain walls coupled with I-\=I solitons. 
After the transition, the broad spectrum has been attenuated, and rather sharp peaks appear instead. 
For the static neutral state, linear excitation spectra are calculated in the RPA and compared with these Fourier spectra.
The solid and dotted lines in Fig.~\ref{fig-fftrpa-in}~(b) show those of the charge-charge correlation and of the current-current correlation, respectively.
Here, the charge density difference between the odd and even sites (wave vector $\pi$) and the total current density (wave vector 0) are respectively used in the RPA calculations.
The peak at energy about $6.2 \times 2 \pi \omega $ in Fig.~\ref{fig-fftrpa-in}~(a) corresponds to the charge-charge correlation.
The position of the lower-energy peak in Fig.~\ref{fig-fftrpa-in}~(a) depends on the number of excited electrons, and it is shifted to the lower-energy side with the increasing number (not shown).
Finally, the position of the lower-energy peak comes close to the peak of the current-current excitation spectrum in Fig.~\ref{fig-fftrpa-in}~(b).
This is interpreted as follows.
After the phase transition, the total electronic energy is still higher than the total energy of the metastable, static and undistorted neutral state.
The excess energy creates many electron-hole pairs in the neutral phase.  
The number of electron-hole pairs increases with the excess energy, and thus with the number of excited electrons.
When the number of electron-hole pairs is large enough, the pairs are overlapped with each other.
These overlaps overcome the somehow disordered lattice displacements so that they enhance the collective nature of electron-hole pairs, i.e., the excitonic nature.
The exitonic effect reduces the energy of charge transfer and slightly slows down the motion of the ionicity.
It is now clear that the rapid oscillation of the ionicity is composed of two modes, one of which is of the excitonic origin.

\subsection{ Dependence on the number of excited electrons}\label{exdep}

\begin{figure}
\includegraphics[height=12cm]{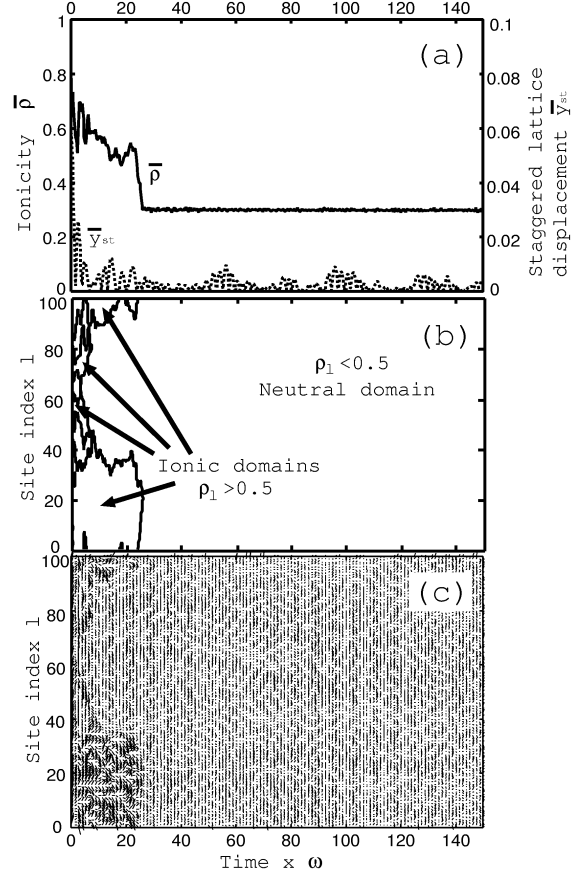}
\caption{(a) Averaged ionicity $\bar{\rho}$ (solid line) and averaged staggered lattice displacement $\bar{y}_{st}$ (dashed line), as a function of time $t$ after the photoexcitations, which is multiplied by the bare phonon energy $\omega$.
(b) Space and time dependence of the boundaries between the neutral and ionic domains ($\rho_l=0.5$).
(c) Correlation between the staggered lattice displacement $y_{st \; l}$ and the ionicity $\rho_l$, as a function of $l$ and $\omega t$.
The parameters are the same as in Fig.~\ref{fig-in}.
In the 100-site chain, 6 electrons are excited.}
\label{fig-ex6}
\end{figure}

We show here the dependence on the number of excited electrons.
When six electrons are excited and the degree of initial lattice disorder is not very small again, $\it{T}$=0.0017 eV, the ionicity and the staggered displacements evolve as shown in Fig.~\ref{fig-ex6}.
The averaged ionicity $\bar{\rho}$ decreases faster [Fig.~\ref{fig-ex6}~(a)] than that after the four-electron excitation [Fig.~\ref{fig-in}~(a)].
Now there are many neutral-ionic domain boundaries [Fig.~\ref{fig-ex6}~(b)].
The neutral domains grow much faster than those after the four-electron excitation, so that the ionic-to-neutral phase transition is completed by the much shorter time about $27/\omega$.
Some I-\=I solitons are seen in Fig.~\ref{fig-ex6}~(c).
The neutral domain grows so rapidly that its motion is insensitive to the I-\=I solitons.

\begin{figure}
\includegraphics[height=12cm]{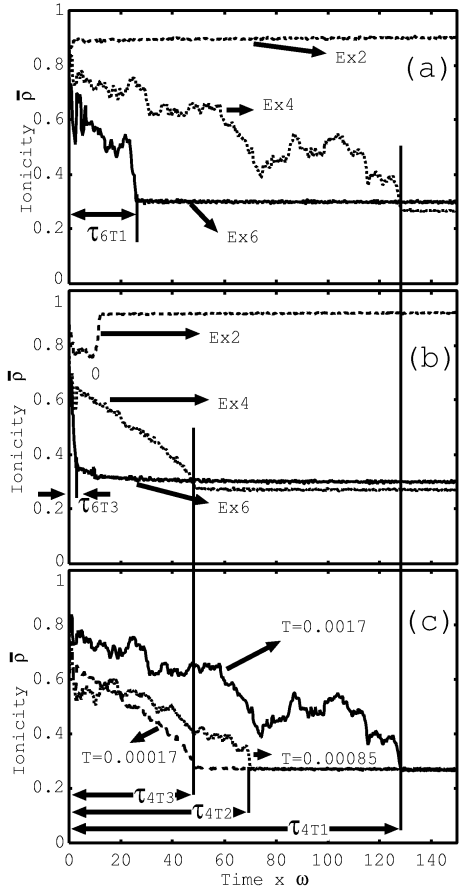}
\caption{Dependence of the evolution of the averaged ionicity $\bar{\rho}$ on the number of excited electrons, (a) with $T=$0.0017 eV, and (b) with $T=$0.00017 eV.
The dashed, dotted, and solid lines show $\bar{\rho}$ after two (denoted by Ex2), four (Ex4), and six (Ex6) electrons are excited, respectively.
(c) shows the dependence on the degree of initial lattice disorder.
The dashed, dotted, and solid lines show $\bar{\rho}$ after 4 electrons are excited with $T=$0.00017 eV, $T=$0.00085 eV, and $T=$0.0017 eV, respectively.
The other parameters are the same as in Figs.~\ref{fig-tl-2ex}, \ref{fig-tl}, \ref{fig-in}, and \ref{fig-ex6}.}
\label{fig-tpdep2}
\end{figure}

The dependence of the evolution of the averaged ionicity $\bar{\rho}$ on the number of excited electrons with the degree of initial lattice disorder, ${\it T}=0.0017$ eV and ${\it T}=0.00017$ eV, is shown in Figs.~\ref{fig-tpdep2}~(a) and \ref{fig-tpdep2}~(b), respectively.
The dashed, dotted, and solid lines show $\bar{\rho}$ after two (denoted by Ex2 in the figures), four (Ex4), and six (Ex6) electrons are excited, respectively, in the 100-site chain.
Here, the time required for the completion of the photoinduced phase transition is denoted by $\tau_{nTm}$, where $n$ describes the number of excited electrons and $m$ distinguishes the degree of initial lattice disorder.
As more electrons are excited, the transition is completed earlier.
It means that the stronger photoexcitations induce the faster ionic-to-neutral transitions.
The dependence on the degree of initial lattice disorder $\it{T}$ is shown in Fig.~\ref{fig-tpdep2}~(c).
The dashed, dotted, and solid lines show $\bar{\rho}$ after four electrons are excited with $T=0.00017$ eV, $T=0.00085$ eV, and $T=0.0017$ eV, respectively.
As $T$ increases, more I-\=I solitons are produced.
Because the I-\=I solitons obstruct the growth of neutral domains, they slow down the transition.
Consequently, the transition period $\tau_{nTm}$ becomes longer as the initial lattice variables are more strongly disordered.
Disorder itself is necessary to nucleate neutral domains.
At $T=0$, neutral domains can be created by numerical errors, but they do not appear immediately, so that the transition is not always completed quickly.
We remark that, in real systems, the free-energy difference and the barrier height between the ionic and neutral phases also depend on the temperature.
As long as the initial randomness in the lattice displacements and velocities are concerned, higher temperatures are unfavorable to reduce the transition period.
However, the other effects remarked here would work on the contrary.

\section{Conclusions}\label{conc}

We have investigated how neutral domains grow or shrink in the ionic background after photoexicitations through their dependence on the excitation intensity and on the degree of lattice disorder in the initial state, using the one-dimensional extended Peierls-Hubbard model.
As the photoexcitation becomes intense, more neutral domains are initially created and their average size increases.
Above threshold intensity, it eventually becomes neutral throughout the system.
After the photoexcitation of the ionic state, neutral domains, \=I domains, and thus I-\=I solitons are created.
The \=I domains (i.e., with the opposite polarization to the initial one) reduce the averaged staggered lattice displacement $\bar{y}_{st}$.
Consequently, the averaged staggered lattice displacement $\bar{y}_{st}$ decays much faster than the averaged ionicity $\bar{\rho}$, in general.
This result is consistent with the experimentally observed, quick decay of the SHG intensity\cite{luty02}, compared with the decrease of the ionicity, although the calculation is performed in the one-dimensional system.
As the initial lattice variables are more strongly disordered, more I-\=I solitons appear.
The I-\=I solitons obstruct the growth of neutral domains.
Therefore, disorder slows down the phase transition.
In short, the progress of the photoinduced ionic-to-neutral phase transition is not simply like dominoes that are falling down, in the sense that the competition between I-\=I solitons and neutral-ionic domain boundaries affect the growth of neutral domains and thus the transition itself.

\section*{Acknowledgement}

We are grateful to H. Cailleau, S. Koshihara, M. H. Lem\'ee-Cailleau, T. Luty and H. Okamoto for fruitful discussions.
This work was supported by the NEDO International Joint Research Grant Program, and a Grant-in-Aid for Scientific Research (C) from Japan Society for the Promotion of Science.



\end{document}